\documentstyle[12pt]{article}

\begin{document}

\vskip 2cm
\begin{center}
 { \Large {\bf ON SPINOR REPRESENTATIONS IN THE WEYL GAUGE THEORY} \\
 \vspace*{1cm}
 {B.M.BARBASHOV and A.B.PESTOV  \\}
\vspace*{1cm}
{\it Bogoliubov Laboratory of Theoretical Physics,\\
 Joint Institute for Nuclear Research, 141980 Dubna, Russia } }
\end{center}
\begin{abstract}
A spinor current-source is found in the Weyl non-Abelian gauge theory
which does not contain the abstract gauge space. It is shown that
the searched spinor representation can be constructed in the space of
external differential forms and it is a 16-component quantity for
which a gauge-invariant Lagrangian is determined.
The connexion between the Weyl non-Abelian gauge potential and
the Cartan torsion
field and the problem of a possible manifestation of the considered
interactions are considered.
\end{abstract}
\section{Introduction}
In Ref. 1 it has been shown that the congruent transference introduced by
Weyl $^2$ in 1921 defines a non-Abelian gauge field. The Weyl gauge theory
is a realization of abstract theory of gauge fields in the framework of
classical differential geometry which does not assume separation between
space-time and a gauge space. At the same time, contemporary gauge
models assume an exact local separation between space-time and a gauge
field. It is just this point at that the Weyl theory opens a new possibility.

It is shown that the space of all covariant antisymmetric tensor fields
is a spinor representation of the Weyl gauge group and allows the
construction of a spinor current-source in  a gauge theory of that type.
Status of the Cartan torsion field within the Weyl gauge theory is
considered from  different points of view.
\section{Gauge potential}
The Weyl connexion which defines the congruent transference of a vector is
of the form
\begin{equation}
{\Gamma}^i_{jk} = \{ ^i_{jk}\} +F^i_{jk} ,
\end{equation}
where $\{^i_{jk}\} $ are components of the Levi-Civita connexion of the
metric $g_{ij}$ usually called the Christoffel symbols:
\begin{equation} \{^i_{jk}\}
=\frac{1}{2} g^{il} (\partial_j g_{kl} + \partial_k g_{jl} - \partial_l
g_{jk}),
\end{equation}
and $F^i_{jk} = F_{jkl} g^{il}$  are components of the Weyl non-Abelian
gauge potential
that is a covariant
third-rank tensor, skew-symmetric in the last two indices
\begin{equation}
F_{jkl} + F_{jlk} = 0.
\end{equation}
According to (1), a vector $v^i$ under the congruent transference
changes by the law
\begin{equation}
dv^i = - \{^i_{jk}\} \,dx^j \,v^k -F^i_{jk} \, dx^j \,v^k ,
\end{equation}
which includes the displacement belonging to the Riemann geometry
and the rotation determined by the metric $g_{ij} $ and the
bivector $F_{jkl}\,dx^j.$ Denote by $\stackrel{w}\nabla_i $ the covariant
derivative with respect to the Weyl
connexion (1). Then with allowance for (3) we obtain
\begin{equation}
\stackrel{w}\nabla_i g_{jk} = 0.
\end{equation}
Thus, the Weyl connexion is metric and this is in agreement with results
obtained  by Hayashi $^3,$  who has found that in macrophysical and
microphysical systems the affine connexion cannot be nonmetric but is
very likely to be metric.

The Weyl geometric construction presented above has a simple group-
theoretical meaning. Let $S^i_j  $ be components of a tensor field $S$ of type
(1,1) obeying the condition $det(S^i_j)\neq 0.$ In this case there exists
a tensor field $S^{-1}$ with components $P^i_j  $ such that
$S^i_k P^k_j = \delta^i_j.$
It is obvious that the tensor field $S$ can be regarded as a linear
transformation
  \begin{equation}
{\bar v}^i = S^i_j \,v^j
\end{equation}
in the space of vector fields; $S^{-1}$ is the inverse transformation. Since
under congruent transference the length of a vector remains constant, among
the transformations (6) we distinguish those that do not change the length
of a vector; they are given by the equations
 \begin{equation}
g_{ik} S^k_j = g_{jk} P^k_i.
\end{equation}
Transformations of the form (6) and (7) form a group that is a gauge
group, as will be shown below; we denote it by
$G_W.$ The gauge group
establishes an equivalence relation in the spaces of different fields. It
can be shown that if a vector $v^i$ in an equivalence class undergoes
congruent transference, then any vector ${\bar v}^i =S^i_j v^j$
equivalent to it in the sense
of the group $G_W,$ also undergoes congruent transference.
In other words, if for a $v^i$ we have (4), then for ${\bar v}^i =S^i_j v^j$
 the formula
$$ d{\bar v}^i = - \{^i_{jk}\} \,dx^j \,{\bar v}^k -{\bar F}^i_{jk} \, dx^j \,{\bar v}^k ,$$
takes place, where
 \begin{equation}
{\bar F}_{lkm} = F_{lij} \,P^i_k \,P^j_m +
g_{ij} \, P^i_k  \, {\nabla_l}P^j_m.
\end{equation}
In (8) and what follows $\nabla_i$ is the covariant derivative with respect
to the Levi-Civita connexion (2).
>From (7) it follows that the tensor ${\bar F}_{lkm}$ obeys equation (3),
and hence, the Weyl connexion
 $${\bar {\Gamma}}^i_{jk} = \{^i_{jk}\} + {\bar F}^i_{jk} $$
is also metric. Thus, with a given metric connexion we have an entire class of
equivalent metric connexions.

Consider an infinitesimal gauge
transformation
$S^i_j = \delta^i_j + u^i_j , \quad P^i_j = \delta^i_j - u^i_j ,$
which upon substitution into (7) gives $g_{ik}u^k_j + g_{jk}u^k_i = 0.$
Hence
it follows that any antisymmetric covariant tensor field of second rank
(2-form) $u_{ij} = - u_{ji}$ determines an infinitesimal gauge
transformation since $ u^i_j = u_{jk}g^{ik}.$
>From (8) we obtain  that an infinitesimal gauge transformations of
the potential have the form
\begin{equation}
{\delta F}_{ijk} =  \nabla_i u_{jk} + F_{ijl} u^l_k -  F_{ikl} u^l_j .
\end{equation} Let us now construct
the strength tensor of the gauge field
\begin{equation}
B_{ijkl} = \nabla_i F_{jkl} -\nabla_j F_{ikl} +
F_{ikm} F_{jl}^m - F_{jkm} F_{il}^m + R_{ijkl},
\end{equation}
where $R_{ijkl} $ is the Riemann curvature tensor of the metric $g_{ij}.$
>From (9)
it follows that the strength tensor is gauge-transformed by the law
$$    {\delta B}_{ijkl} =  B_{ijkm} u^m_l -  B_{ijlm}u^m_k.$$
Let us interpret the Riemann curvature tensor in expression (10) from a
group-theoretical and geometric point of view. We set
$B_{ijkl} = H_{ijkl} + R_{ijkl} $ and from
(9) and (10) we get
$${\delta H}_{ijkl} =  H_{ijkm}u^m_l - H_{ijlm}u^m_k + (\nabla_i
\nabla_j - \nabla_j\nabla_i)u_{kl}. $$
According to the Ricci identities, we find
$$(\nabla_i\nabla_j - \nabla_j\nabla_i) u_{kl} = R_{ijkm}u^m_l -
R_{ijlm}u^m_k,$$
which clearly shows the role of the Riemann curvature tensor under gauge
transformations. The tensor (10) has a simple geometric meaning. It can
be shown that the curvature tensor of the Weyl connexion (1) coincides with
the strength tensor (10) of the Weyl non-Abelian gauge field whereas
the gauge potential is
considered as a deformation tensor of the Levi-Civita connexion (2).

Thus the tensor field $F_{ijk}, $ entering into the Weyl connexion
is a gauge
field, and the tensor $B_{ijkl}$ is the strength tensor of that field. We
stress that the gauge group in the case under consideration is defined
by the metric, while the gauge field has a direct geometrical meaning
(congruent transference) and no extra internal or gauge space is to be
introduced. Here gauge symmetry reflects the fact that there does not
exist any objective property that could distinguish the geometry defined
by the connexion ${\Gamma}^i_{jk} = \{ ^i_{jk}\} +F^i_{jk} $ from the
one defined by the connexion
${\bar \Gamma}^i_{jk} = \{ ^i_{jk}\} + {\bar F}^i_{jk} .$

\section{Gauge-field equations}
We write the gauge-invariant Lagrangian in the form
\begin{equation}
      L = -\frac{1}{16} B_{ijkl} B^{ijkl} + \frac{1}{4} F_{ijk}S^{ijk},
\end{equation}
where $S^{ijk} $ is an unknown current-source of the gauge field that should
be a quadratic function of components of the quantity defining a spinor
representation of the gauge group $G_W.$ Variational procedure results in
the following equations of the gauge field:
\begin{equation}
\nabla_i B^{ijkl}+ F^k_{im} B^{ijml} - F^l_{im} B^{ijmk}+ S^{jkl} = 0.
\end{equation}
>From these equations we derive the equations for the gauge-field
current-source
\begin{equation} \nabla_{i} S^{ikl} + F^k_{im} S^{iml} - F^l_{im}
S^{imk} = 0.
\end{equation}
Next, consider the current vector
$$ Q^j =
\frac{1}{2}v_{kl}( F^k_{im} B^{ijml} - F^l_{im} B^{ijmk}+ S^{jkl}). $$
>From the field equations it follows that the current is conserved if the
bivector $v_{ij}$ obeys the equation $ \nabla_{i} v_{jk}= 0.$
However, the corresponding conserved quantity is not gauge-invariant. The
same holds true also in the abstract theory of gauge fields. In all the
previous formulas it was assumed that the gauge potential is of dimension
of the inverse length. To introduce the constant of interaction with the gauge
field, we should make the substitution
$F_{ikl} \Rightarrow ({\varepsilon}/{\hbar c}) F_{ikl}.$
In the limit $\varepsilon \Rightarrow 0 $ the
Lagrangian (11) transforms into the pure gravitational one
$$L = - \frac{1}{16} R_{ijkl} R^{ijkl},$$
which is known $^4$ to be renormalizable.

Let us now compare the Weyl non-Abelian gauge theory with the
abstract theory of gauge fields. The latter is based on an arbitrary
semisimple Lie group with structure constants $f^a_{bc} $ and a set
of vector fields. Space-time indices are raised and lowered with the
metric tensor $g_{ij}$, whereas parametric indices, with the group
metric $^5$; $g_{ab} = f_{am}^n f_{bn}^m.$ In the Weyl
non-Abelian gauge theory, the metric tensor is also a group tensor
and structure constants are absent. The reason is that for some Lie
groups, and for the group in question as well, the coordinates on a
group can be regarded as tensor fields in space-time, which just
leads to the situation when space-time and gauge space are not
separated like in the abstract theory.

\section{Spinor representation}
Let us consider the field that is a source of the Weyl non-Abelian
gauge field and
defines a spinor representation of the group $G_W.$
Consider a 16-component object which can
be defined as space of all covariant antisymmetric tensor fields
$f_{i_1\cdots i_p} (p= 0,1,2,3,4)$
on a space-time manifold with the metric $g_{ij}.$
Mathematically, a shorted notation 'differential form' is adopted $^6.$
So, the form is the following quantity
\begin{equation}
F = (f, \quad f_i, \quad f_{ij}, \quad f_{ijk}, \quad f_{ijkl}).
\end{equation}
Objects of that sort were first considered in Ref.7 (see also Refs.8 - 11).

The spinor representation of the Weyl gauge group is the field of type (14).
To prove this statement, we determine the natural Lagrangian for the
field (14) and show that it is invariant under gauge transformations
which define the symmetry aspect of the Weyl non-Abelian gauge field.
We define the
scalar bracket of two fields of the type (14) as follows
$$(F,H) = \bar f h + \bar f_i h^i + \frac{1}{2!}\bar f_{ij} h^{ij} +
\frac{1}{3!}\bar f_{ijk} h^{ijk} + \frac{1}{4!}\bar f_{ijkl} h^{ijkl},$$
where the bar means complex conjugation. If $F$ is a form, the generalized
curl operator $d$ is given as follows
\begin{equation}
dF = (0, \, \partial_i f, \, 2\partial_{[i} f_{j]}, \,
3\partial_{[i} f_{jk]}, \, 4\partial_{[i} f_{jkl]}).
\end{equation}
Here square brackets denote alternation;
$\partial_i = \partial / \partial x^i. $ The simplest Lagrangian
for the field $F$ that can be constructed in terms of the operator $d$ is
of the form
\begin{equation}
L_d(F)= (F,dF) + (dF,F) + m(F,F),
\end{equation}
where $m$ is the mass of a particle $ ( c = \hbar = 1).$
Note that the operator of external differentiation (15) is the only
linear operator of first order that commutes with transformations of the
group of diffeomorphisms, the group of symmetry of gravitational
interactions. Therefore, the Lagrangian (16) is defined uniquely. If
$ \nabla_i $ is a covariant derivative with respect to the
Levi-Civita connexion of the metric $g_{ij},$ defined by relations
(2), then partial derivatives in (15) can be replaced by covariant
derivatives. The Lagrangian (16) is not suitable for the
investigation since the operator $d$ is not self-conjugate with
respect to the scalar product $$<F|H> = \int (F,H)\sqrt{ -g} d^4x.$$
Using the identity
$$\sum_{p=0}^4 \frac{1}{p!} f_{k i_1\cdots
i_p} h^{k i_1\cdots i_p}= \sum_{p=0}^4 \frac{1}{p!} (p f_{i_1\cdots
i_p}) h^{i_1\cdots i_p}$$
we can easily verify that the operator $\nabla = \delta + d,$
possesses the required
property, where $\delta$ is the operator of generalized divergence
$$\delta F = (-\nabla^m f_m, \, -\nabla^m
f_{mi}, \, -\nabla^m f_{mij}, \, -\nabla^m f_{mijk}, \, 0) .$$
The Lagrangian (16) in terms of the operator $\nabla = \delta + d$ reads
$$L_d(F) = \frac{1}{2}(F, \nabla F) + \frac{1}{2}( \nabla F, F) +
m (F,F) + \nabla_i T^i,$$
where
$$T^k = \sum_{p=0}^4 \frac{1}{p!}( \bar
f_{ i_1\cdots i_p} f^{k i_1\cdots i_p} + \bar f^{k i_1\cdots i_p} f_{
i_1\cdots i_p}) . $$
So, the Lagrangian (16) is equivalent to the Lagrangian
\begin{equation}
L(F) = \frac{1}{2}(F, \nabla F) +
\frac{1}{2}( \nabla F, F) + m (F,F) ,
\end{equation}
which will be now analyzed. We define a numerical operator $\Lambda,$ by
setting
$$\Lambda F = (f, \, - f_i, \, f_{ij}, \, -f_{ijk}, \, f_{ijkl}) .$$
It is not difficult to verify the validity of the following relations
\begin{equation}\Lambda^2 = 1, \quad \Lambda d + d \Lambda = 0, \quad
\nabla \Lambda + \Lambda \nabla = 0.  \end{equation}
Since $\nabla d + d \nabla = \nabla^2,$ then we have
\begin{equation}
\nabla (\frac{1}{2}\nabla - d) + (\frac{1}{2} \nabla - d) \nabla = 0.
\end{equation}
>From (18) and (19) it follows that the operator
\begin{equation} \stackrel{*}{\nabla}
= (\nabla - 2d) \Lambda \end{equation}
commutes with the operator $\nabla,$ whereas their squares are equal
 $$\nabla
\stackrel{*}{\nabla} = \stackrel{*}{\nabla} \nabla,  \quad \nabla^2 =
\stackrel{*}{\nabla}{^2}.$$
We will call the operator
$\stackrel{*}{\nabla}$ dual to the operator $\nabla.$ This duality property of
the field (14) allows us to introduce the important operators
in the following way.
In accordance
with the principle of 'minimal electromagnetic interaction', we make
the substitution $\nabla_i \Rightarrow \nabla_i - \frac{ie}{\hbar c} A_i$
in the operators $\nabla$ and $\stackrel{*}{\nabla},$
denote the new
operators by $D$ and $\stackrel{*}{D}$, respectively, and determine their
squares. We have
 $$\stackrel{*}{D}{^2} = \nabla^2 -
\frac{ie}{\hbar c} Q(F_{ij}) + \frac{2ie}{\hbar c} A^i \nabla_i +
\frac{e^2}{\hbar^2 c^2} A_i A^i + \frac{ie}{\hbar c} \nabla_i A^i,$$
where $F_{ij},$ is a bivector of the electromagnetic field. A similar formula
follows for the dual operator $D$ with the change of the operator
$Q(F_{ij}) $ by the dual
operator $\stackrel{*}{Q}(F_{ij}).$ The operators $Q(F_{ij}) $ and
$\stackrel{*}{Q}(F_{ij})$
are defined by antisymmetric tensor
fields of second rank (2-forms). Let us write the operators
  $Q(u_{ij}), \quad
 \stackrel{*}{Q}(u_{ij})$
in an explicit form
\begin{eqnarray} Q(u_{ij}) F & = &
 (\frac{1}{2} u^{mn}f_{mn}, \, \frac{1}{2} u^{mn}f_{mni} + u_{mi} f^m,\nonumber\\ & &
 \, \frac{1}{2} u^{mn}f_{mnij}+ 2u_{m[i} f^m_{.j]} - u_{ij}f,
 \nonumber\\ & & 3u_{m[i}f^m_{.jk]} - 3 u_{[ij} f_{k]},  \,
  - 6u_{[ij} f_{kl]} ),  \end{eqnarray}
\begin{eqnarray} \stackrel{*}{Q}(u_{ij}) F & = &
 (- \frac{1}{2} u^{mn}f_{mn}, \, -\frac{1}{2} u^{mn}f_{mni} + u_{mi}
 f^m,\nonumber\\ & & \, -\frac{1}{2} u^{mn}f_{mnij}+ 2u_{m[i}
 f^m_{.j]} + u_{ij}f, \nonumber\\ & & 3u_{m[i}f^m_{.jk]} + 3 u_{[ij}
 f_{k]},  \,  6u_{[ij} f_{kl]} ). \nonumber \end{eqnarray}

It can be shown that the operators $Q(u_{ij}) $ and $\stackrel{*}Q(u_{ij})$
commute and this is another manifestation of the duality. Algebra of the
operators $J(u_{ij}) = \frac{1}{2}Q(u_{ij})$ is closed with respect to the
Lie bracket operation, i.e.
\begin{equation}
[J(u_{ij}), \, J(v_{ij})] = J(w_{ij}),
 \end{equation}
where
\begin{equation} w_{ij} = u_{im} v_{j.}^m - u_{jm}
 v_{i.}^m.
\end{equation}
>From (23) it follows that the operators $J(u_{ij})$ define a representation
of the considered Weyl group $G_W$ in the space of the fields (14). Since
\begin{equation}
(F, J(u_{ij}) H) = - (J(u_{ij})F, H),
\end{equation}
then the Lagrangian (17) will be invariant under the gauge transformations
 \begin{equation} F \Rightarrow \bar F = exp (J(u_{ij})) F,
\end{equation}
provided that
\begin{equation}
[J(u_{ij}), \quad \nabla] = 0.
\end{equation}
The relation (26) holds valid if the bivector $u_{ij} $ satisfies the equations
\begin{equation}
\nabla_i u_{jk} = 0.
\end{equation}
The conditions of integrability of equations (27) follow from the Ricci
identities and are of the form
$R_{ijk.}{}^m u_{ml} + R_{ijl.}{}^m u_{km} = 0.$  When
$ R_{ijl.}{}^m = K(g_{il}\delta^m_j - g_{jl}\delta^m_i ),$ equations
(27) will not have
solutions at all. Thus, the Lagrangian (17) in the space of constant
curvature will be invariant under the transformations (25) only upon
introducing a gauge field of a definite type. The latter can be determined
as follows. Consider variations of the type $\delta F = J(u_{ij}) F.$
This class of variations,
up to the Lagrange derivative, yields for the Lagrangian (17)
 $$\delta
L(F) = \frac{1}{4} \nabla_i u_{jk}S^{ijk},$$
where $S^{ijk} $ is a tensor field of third rank antisymmetric in the last two
indices
\begin{eqnarray} S^{jkl} & = & \sum_{p=0}^4 \frac{1}{p!} (\bar
f^j_{.i_1\cdots i_p} f^{kli_1\cdots i_p} + 2 g^{j[k}\bar f^{l]}_{.i_1
\cdots i_p} f^{i_1\cdots i_p} - \nonumber\\ & & - 2\bar f^{i_1\cdots
i_p j[k} f^{l]}_{.i_1\cdots i_p} - \bar f_{i_1 \cdots i_p}
 f^{jkli_1\cdots i_p})+ c.c..  \end{eqnarray}
So, the Lagrangian (17) is to be supplemented with a term of the form
$$L_I = \frac{1}{4}F_{jkl} S^{jkl}$$
to ensure gauge invariance. We added the same term to the Lagrangian (11)
of the gauge field. Thus, the explicit form of the current source of the
gauge field is determined uniquely. From (28) it follows that under
transformations $\delta F = J(u_{ij})F$, the tensor $S^{jkl}$ is transformed
by the law
$$ \delta S^{jkl} = u_{m}^k S^{ijm} - u_{m}^j S^{ikm}.$$
Hence we obtain that the gauge field $F_{ijk}$ is transformed as follows:
$$\delta F_{ijk} =  \nabla_i
u_{jk} + F_{ijm} u_{k.}^m - F_{ikm} u_{j.}^m.$$
According to (9), the field $F_{ijk}$ is the Weyl non-Abelian gauge
field, whereas the field $F$ is shown to be its spinor source. That
the transformations (25) define the spinor representation of the
group $G_W$ can easily be verified by comparing them with the
transformations (6). Let $a_i$ and $b_i$ be the unit orthogonal
covectors, $(a,a) = (b,b) = 1, \quad (a,b)= 0,$  where $(a,b) = a_i
b_j g^{ij},$  and $u_{ij} = \alpha (a_i b_j - a_j b_i), \quad (u =
\alpha a\wedge b).$ If $Rv^i = (a_j b^i - a^i b_j)v^j ,$ then after
some calculations we have $$\{exp(\alpha R)\}v^i =  v^i - a^i
(a,v) - b^i (b,v) + $$ $$ + \{ ({\cos \alpha}) a^i + ({\sin \alpha})
b^i\}(a,v) + \{({\cos \alpha}) b^i - ({\sin \alpha}) a^i\} (b,v).$$
For the operator $J(u_{ij})$ we have $J^2 = - (1/4) \alpha^2$  and
hence $$exp(\alpha J(a\wedge b)) = \cos\frac{\alpha}{2} +
(\sin\frac{\alpha}{2}) J(a\wedge b).$$  By setting $\alpha = 0, \,
2\pi,$ it is not difficult to verify that the vector fields are a
tensor representation of the group $G_W$ whereas the space of fields
(14) is the carrier space of a spinor representation of the gauge
group in question.

Theory of the field $F_{ijk}$
has been already formulated above, and in the next section we dwell upon
the relation between the Weyl gauge potential and the Cartan torsion.
That this relation does exist follows from both the fields being tensor fields
of the same type.

\section{Torsion and gauge symmetry}
At present, the torsion discovered by Cartan is the subject of numerous
studies aimed at establishing its physical meaning  and the connexion of
general relativity with the physics of microworld. We will consider this
question in the framework of the Weyl non-Abelian gauge theory. Let an
affine connexion
be given, and $\Gamma^i_{jk}$ be its components.
Then, as it
was first shown by Cartan $^{12}$, the affine connexion uniquely defines a
tensor field  $T^i_{jk} = (\Gamma^i_{jk} - \Gamma^i_{kj})/ 2,$ that is
called the torsion tensor.
The Riemann-Cartan space-time $U_4$ is a paracompact, Hausdorff, connected
$C^{\infty}$ 4-dimentional manifold endowed with a locally Lorentzian
metric $g_{ij}$ and an affine connexion $\Gamma^i_{jk}$  which is
metric
$$ \partial_i g_{jk} - \Gamma^l_{ij} g_{lk} -\Gamma^l_{ik} g_{jl} =0.$$
The Riemann-Cartan space-time has both the curvature tensor and the torsion
tensor. In terms of the torsion tensor the solution of the last equation
may be represented in the form
\begin{equation}
\Gamma^i_{jk} =
\{^i_{jk} \} + T^i_{jk} + g^{il} T^m_{lj} g_{mk} + g^{il} T^m_{lk}g_{mj}.
\end{equation}
Thus, the connexion of the Riemann-Cartan space-time
is defined unambiguously.

 We will show that the Cartan torsion cannot be introduced in the framework
of Weyl  non-Abelian gauge theory in a gauge-covariant manner.
As it follows from (1), the torsion tensor of the Weyl connexion is equal to
\begin{equation}
  T^i_{jk} =  (F^i_{jk} - F^i_{kj})/2
\end{equation}
or
\begin{equation}
T_{ijk}= (F_{ijk} - F_{jik} )/2,
\end{equation}
where $T_{ijk} = T^l_{ij} g_{lk}.$
Similarly the (31), for the Weyl connexion
${\bar {\Gamma}}^i_{jk} = \{^i_{jk}\} + {\bar F}^i_{jk},$  equivalent (1),
we have
\begin{equation}
\bar T_{ijk}=  (\bar F_{ijk} - \bar F_{jik})/2 .
\end{equation}
In view of (8)
it is entirely natural to pose the question on the relationship between the
tensors  $\bar T_{ijk}$ and $T_{ijk}.$  However, such a relationship that
contains only the tensors $\bar T_{ijk},$  $T_{ijk}$ and the elements
of the gauge group does not exist. It is easy to see from the
formulas
$${\bar F}_{ijk} =
F_{ilm} \,P^l_j \,P^m_k + g_{lm} \, P^l_j \,{\nabla_i}P^m_k, $$
$${\bar F}_{jik} = F_{jlm} \,P^l_i \,P^m_k +
      g_{lm} \, P^l_i  \, {\nabla_j}P^m_k. $$
Indeed, since the tensor $F_{ijk}$
is a skew-symmetric with respect to the second and third indices, whereas
the torsion tensor is skew-symmetric with respect to the first two indices,
in the relation (31) the index that participates in the gauge transformation
(8) and an index that is not affected by it are confused. Now it is clear
why the torsion determines the  metric connexion uniquely.

The conclusion is that the torsion tensor is not a geometrical quantity from
the point of view of gauge symmetry. The tensor $T^i_{jk}$ does not define a
representation of the gauge group. It may be said that the concept of the
torsion tensor is not gauge-covariant.
Specifying the torsion tensor, we definitely fix
the gauge.  Thus, from the point of view a symmetry,
the fundamental geometrical object is the tensor $F_{ijk}$
that determines the congruent transport. It is for this tensor that the
gauge-invariant equations (12)  are written down,  which are in fact
determined uniquely by the gauge symmetry.  It is now easy to
understand why for the torsion tensor all possible Lagrangians are
encountered and investigated in literature with equal success. If one
does pose the question of equations for the torsion, then it is most
natural to  this end to fix the gauge in accordance with what was
said earlier.

We note an interesting connexion between gauge transformations and Riemannian
geometry. The second term on the right-hand side of relation (8) vanishes
if $\nabla_{l} P^i_{m} = 0.$ In the standard theory of gauge fields,
this corresponds to transition from local to global transformations.
In the considered case, the equations $\nabla_{l} P^i_{m} = 0 $ may
not have any  nontrivial solutions at all, for example, in the case
when $g_{ij}$ is the metric of a space of constant curvature. Thus, a
Riemannian geometry in general requires a local (gauge) symmetry. We
note also that geometrical relationships, like physical laws, depend
neither on the choice of the coordinate system nor on the choice of
the basis in the studied vector spaces, so that all the relations
that have been established above can be expressed in any coordinate
system and in any basis, including an orthogonal one.

\section{Field equations and Riemann-Cartan geometry}
In this section we give an explicit example of the relation of
the Weyl non-Abelian gauge field to the torsion tensor in definite gauge
that is defined as follows.
We take the gauge-invariant Lagrangian
\begin{equation}
L = L(F) + \frac{1}{4} F_{ijk} S^{ijk}
\end{equation}
and determine its variation with respect to $F.$
As a result, we have
\begin{eqnarray} \delta L & = & \delta L(F) +
\frac{1}{2}\sum_{p=0}^4 \frac{1}{p!} \delta \bar f^{i_1\cdots i_p} (- F^m
f_{mi_1\cdots i_p} + p F_{[i_1} f_{i_2\cdots i_p]} + \nonumber\\ & &
+\frac{p}{2} D_{mn[i_1} f^{mn}_{..i_2\cdots i_p]} + \frac{p(p-1)}{2}
D_{[i_1i_2|m|} f^m_{i_3\cdots i_p]} - \nonumber\\ & & - \frac{1}{3!}
C^{jkl} f_{jkli_1\cdots i_p} - \frac{1}{3!}p(p-1)(p-2) C_{[i_1i_2i_3}
f_{i_4\cdots i_p]}) + \nonumber\\ & & c.c., \end{eqnarray}
where $F_m = g^{jk} F_{jkm},$
$$C_{ijk} = 3 F_{[ijk]}, \quad D_{ijk} = -C_{ijk} +2 F_{ijk}.$$
Indices sandwiched between vertical lines are not subject to the operation
of alternation.

Next we replace
the covariant derivative $\nabla_i$ with respect to the
Levi-Civita connexion in the Lagrangian $L(F)$
by the covariant derivative $\stackrel{*}{\nabla}_i$ with respect to the
connexion (29) of
the Riemann-Cartan space-time. This peculiar substitution introduces
the torsion field into the Lagrangian (17). A new Lagrangian will be
denoted by $L_{K}.$ According to (29), this Lagrangian for the field
$F$ in the Riemann-Cartan space can be represented as a sum of the
Lagrangian (17) and an extra term to be denoted as $L_{A}(F),$
$L_{T} = L(F) + L_{A}(F). $  With this notes we vary the Lagrangian
$L_{T}$ with respect to $F$  and get \begin{eqnarray} \delta L_{T}(F)
& = & \delta L(F) + \sum_{p=0}^4 \frac{1}{p!} \delta \bar
f^{i_1\cdots i_p} (T^m f_{mi_1\cdots i_p} - p T_{[i_1} f_{i_2\cdots
i_p]} - \nonumber\\ & & -pT_{[i_1|mn|} f^{mn}_{..i_2\cdots i_p]} -
p(p-1)T_{m[i_1i_2} f^m_{i_3\cdots i_p]}) + c.c.,  \end{eqnarray}
where $T_i = T^m_{mi}$  is the covector of torsion.
When varying the Lagrangian $L_{T},$ we
should take into account that
   $$\stackrel{*}{\nabla}_i A^i = \frac{1}{\sqrt
     {-g}}\partial_i(\sqrt{-g} A^i) + 2 T_i A^i,$$
where $g$ is the determinant of the metric tensor.
Now we raise the question about the connexion of the Eiler-Lagrange equations
for the field $F$ that follows from the lagrangians
 $L = L(F) + \frac{1}{4} F_{ijk} S^{ijk}$ and
  $L_{T} = L(F) + L_{A}(F). $
>From comparison of (34)
and (35) it can be seen that these expressions will coincide if the gauge
condition $C_{ijk} = 0$ is imposed on the field $F_{ijk},$ i.e. if we set
 $$ F_{ijk} + F_{jki}+ F_{kij} = 0$$
and then set that  $ F_{ijk} = -2 T_{jki}.$

Thus, we have shown that the field-$F$ equations derived by varying the
Lagrangian (33) can, in a certain gauge, be represented as equations in
the Riemann-Cartan space with the following constraint on the torsion tensor
 $$ T_{ijk} + T_{jki}+ T_{kij} = 0.$$
We write the equations for the field $F$  in the Riemann-Cartan space-time
$$
- {\hat \nabla}^l f_{li_1\cdots i_p} + p
{\hat \nabla}_{[i_1} f_{i_2\cdots i_p]} + m f_{i_1\cdots i_p} = 0, $$
where ${\hat \nabla}_i = \stackrel{*}{\nabla}_i - T_i, $ and $p = 0,1,2,3,4.$
These equations coincide with the gauge invariant equations for the field
$F$  under the conditions  described above.

\section{Conclusions}

We summarize the obtained results and present some problems. The
interpretation of congruent transport given here makes it possible to
establish a deep
connexion between classical differential geometry and
the theory of gauge fields. It is important to emphasize once more the
fundamental significance of this relationship, which is that in the considered
case it is not necessary to introduce an abstract gauge space. The equations
for interacting fields can in fact be uniquely derived. The relations
established for the Weyl non-Abelian gauge field and the Cartan torsion
make it possible
to consider, from a new point of view, the problem of physical interpretation
of the torsion  in the framework of the gauge principle.

The existence of the spinor source of the Weyl gauge field is an interesting
feature of this field that dictates the question about  possible physical
manifestations of this kind of interactions. In the Minkowski space-time
equations (27) are quite integrable. Thus, the gauge symmetry can be
considered in this case as a global one. With respect to this global symmetry
a space of forms (14) is reducible. Associated reduction of the space of forms
(14) gives the Dirac theory
in which we find only well known interactions.
In contrast with this case, there is a more interesting possibility, when
equations (27) have no solutions at all. As it was mentioned above,
this situation occurs in the space of constant curvature,
where the appearance of the
Weyl non-Abelian gauge field in a definite sense becomes  necessary
because of the full
absence of global internal symmetry. A very interesting space-time
of this kind is the de Sitter one which is usually considered
as a cosmological model.
So, the Weyl non-Abelian forces could be manifested on the cosmological
scale. Of course, this do not close the realm of microphysics.
The general remark is that all questions and problems discussed
in literature in relation to the physical interpretation of torsion
can be investigated
in a more suitable framework of the Weyl non-Abelian gauge theory.
\vskip 1cm
\begin{flushleft}
{\bf References}
\end{flushleft}
\begin{enumerate}
\item B.M.Barbashov and A.B.Pestov, {\it Mod. Phys. Lett.}
{\bf A10}, 193 (1995).
\item H. Weyl, {\it Space-Time-Matter} \, (Dover
Publications:  INC, 1922).
\item K. Hayashi, {\it Phys.Lett.} {\bf 65B}, 437 (1976).
\item K.S. Stelle, {\it Phys.Rev.} {\bf D16}, 953 (1977).
\item R. Utiyama, {\it Progr. Theor. Phys.} {\bf 101}, 1596 (1956).
\item T. Eguchi, P.B. Gilkey and A.J.Hanson,
{\it Phys.Rep.} {\bf 66}, 213 (1980).
\item D.Ivanenko and L.Landau, {\it Zeitschrift fur Physik} {\bf 48},
 340 (1928).
\item P. Rashevskii, {\it The theory of spinors; Amer. Math. Soc.,
Translations, Series 2 } {\bf 6}, 1 (1957).
\item E. K\"ahler, {\it Rend. Mat. (3-4)} {\bf 21}, 425 (1962).
\item A.B. Pestov, {\it Theor. and Math. Phys.} {\bf 34}, 48 (1978).
\item J.A.Bullinaria, {\it Ann. of Phys.} {\bf 168}, 301 (1986).
\item \'E. Cartan, {\it Comptes Rendus de l`Academie des Sciences
(Paris)} {\bf 174}, 593 (1922).
\end{enumerate}
\end{document}